# STUDY OF PERIPHERAL COLLISIONS OF $^{56}$FE AT 1 A GeV/c IN NUCLEAR EMULSION

E. FIRU[1], V. BRADNOVA[2], M. HAIDUC[1], A. D. KOVALENKO[2], A. I. MALAKHOV[2], A. T. NEAGU[1], P. A. RUKOYATKIN[2], V. V. RUSAKOVA[2], S. VOKAL[2], P. I. ZARUBIN[2]

*[1] Institute for Space Sciences, P.O. Box MG-23, 76-900 Bucharest-Maguerele, Romania, E-mail: elena.firu@spacescience.ro*
*[2] Joint Institute for Nuclear Research, Dubna, Moscow oblast, 141980, Russia*

*Abstract.* We have investigated the interactions of 1 A GeV $^{56}$Fe in nuclear emulsion. We measured the charge and the angular distributions of single and multiple charged relativistic particles emitted from peripheral interactions. We investigate a possible new method of separating interactions of electromagnetic origin. We provide the values of several parameters evaluated in a sample selected using the classical method, and the method proposed in this paper.

## 1. INTRODUCTION

The experimental data exploited in this work is part of a vast effort devoted to the collection of nuclear data and to the investigation of the nuclear structure of nuclei and their properties, within the framework of the Becquerel [1], collaboration. It also develops methodical challenges in the use of nuclear emulsions.

For many years the nuclear emulsion technique has been used to investigate hadron-nucleus and nucleus-nucleus interactions. Nuclear emulsion detectors have some unique advantages, e.g. 4π detection, sub-micron resolution, excellent detections of both relativistic and very low energy particles, and last but not least, simplicity and relatively low costs in comparison with other techniques. One of the main disadvantages is the composite nature of the target which basically consists of three groups of nuclei: hydrogen, light (carbon, nitrogen, oxygen) and heavy nuclei (bromine, silver).

In the present work we investigate the interaction of 1 A GeV iron nuclei in emulsion and compare it with other investigations [2, 3, 4]. The main purpose of the other works mentioned above was the study of the general properties of the interactions, central collisions and phase transition, unusual properties of nuclear matter under conditions of high pressure and high temperatures. Our purpose is to select mainly peripheral interactions and to study their properties.

## 2. EXPERIMENT

A stack of BR-2 nuclear research emulsion pellicles, 550 $\mu$m thick, sized 10 x 10 $cm^2$ was exposed to a $^{56}Fe$ beam parallel to the emulsion surface at 1 GeV/nucleon at JINR Nuclotron. Each beam trajectory was scanned along the track under high magnification (100 x 1.5 x 10) $\mu$m in order to obtain a sample with minimum detection bias, starting at a distance of 2 cm from the entrance. The tracks of individual nuclei were readily traced from the point of entry to an interaction, or to the point of exit from the microscope travel. A total length of about 28.4414 m of the primary $^{56}F$ beam tracks was scanned. In this manner 383 interactions were located and then analysed in detail. We found the value for the interaction mean free path $\lambda$=7.53 $\pm$ 0.14 cm which is in reasonable agreement with the values found in other experiments (Table 1).

| | Energy (A GeV/c) | Mean Free Path (cm) | Ref. |
|---|---|---|---|
| $^{56}Fe$ | 1.7 | $8.4 \pm 0.2$ | [2] |
| $^{56}Fe$ | 1.8 | $7.63 \pm 0.21$ | [3] |
| $^{56}Fe$ | 1.7 | $7.97 \pm 0.19$ | [4] |
| $^{56}Fe$ | 1 | $7.53 \pm 0.14$ | [present work] |

Table 1 - Mean Free Path from different experiments in nuclear emulsion

In a non-homogeneous target-detector such as nuclear emulsion, one measures the reaction mean free path rather than cross section. Nevertheless we calculate the cross section value from formula $\sigma = 1/N_t\lambda$ (where $N_t$ = $7.967 \times 10^{22}\, atoms/cc$ is the concentration of $A_T$ nuclei in emulsion and $\lambda$ is the experimental mean free path for nuclear interaction). Using the chemical composition of nuclear emulsion described in [5] we found that the cross section value is 1666.9 mb. This can also be calculated using the Bradt-Peters relation [6]:

$$\sigma = \pi[r_0(A_B^{1/2} + A_T^{1/2} - b)]^2$$

where b represent the overlap parameter, $A_B$ corresponds to the projectile mass and $A_T$ corresponds to target mass and $r_0$ is the constant in the following expression

$$r = r_0 A^{1/3}$$

where r is the nuclear radii.

A parametrisation [7] based on data from a variety of emulsion experiments with different projectiles yields $b = 1.16 \pm 0.06$ and $r_0 = 1.25 \pm 0.01 fm$. ($\langle A \rangle = \sum N_i A_i / N$ ) [8].

Using these parameters we obtain $\sigma = 1623$ mb, a value which is in very good agreement with the experimental value.

We found 383 Fe-Em interactions which were used for further analysis. Charged secondaries in these events were classified into the following types: "black" particles (b-particles) with a range in emulsion R$\leq$ 3 mm ($E_{protons} \leq 26$ MeV), "grey" particles (g-particles) with R$\geq$ 3 mm and velocity $\beta = v/c < 0.7$ ($26 < E_{protons} < 375$), (for b and g-particles together one uses the term "heavy-track producing particles" h-particles), shower particles (s-particles) - singly charged particles with $\beta > 0.7$, and "non-interacting" fragments of the projectile nucleus (f-particles) with charge Z ≥ 2.

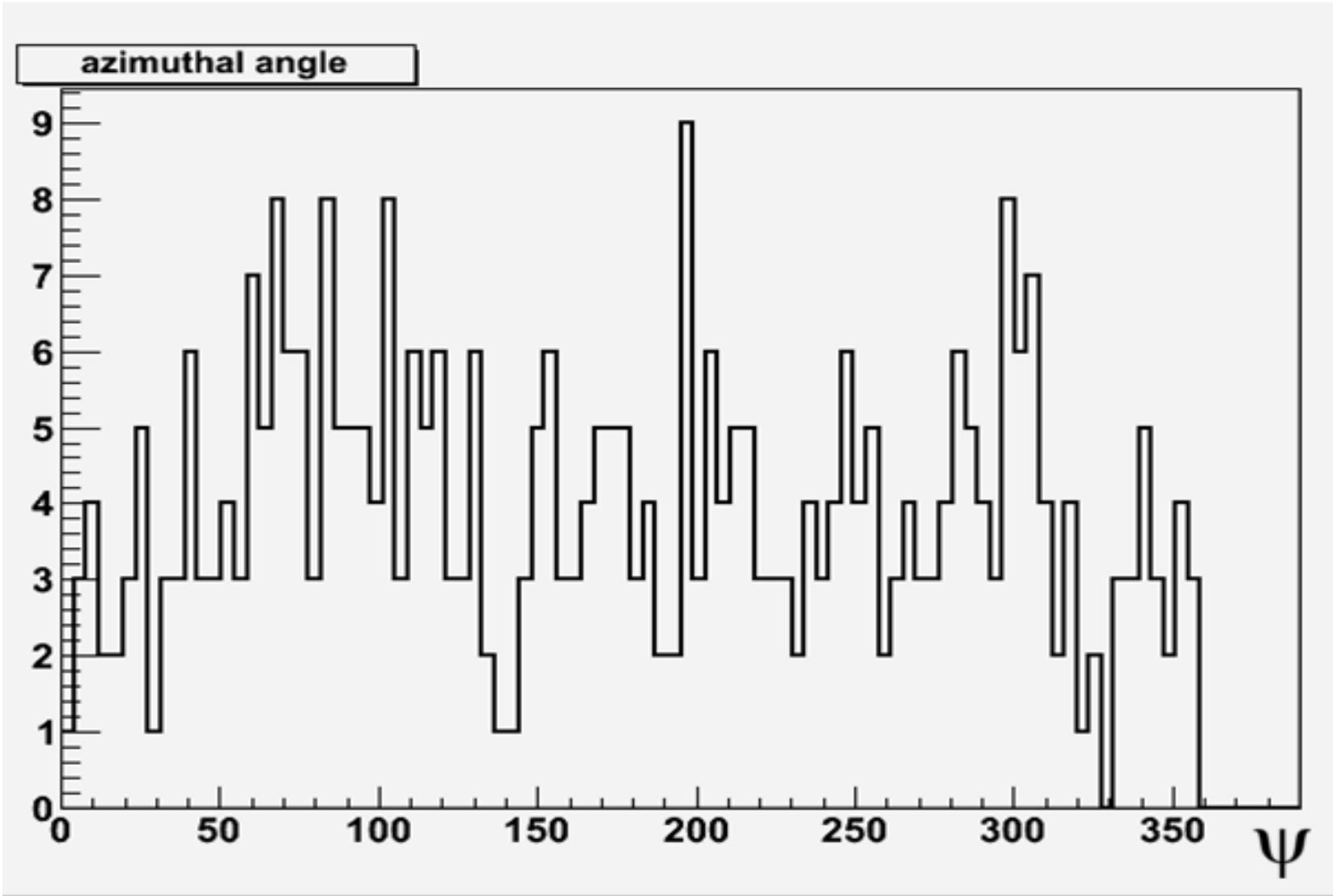

Fig. [1] - Distribution of azimuthal angles of all measured events

Selection of f-particles in nuclear emulsions is easy to perform since these tracks are characterized by the presence of delta rays and high ionization which, in contrast to b and g-particles, does not change over large distances (several cm). Furthermore they do not reveal noticeable multiple scatterings.

The polar $\theta$ and azimuthal $\psi$ emission angles were measured by the coordinate method using a semi-automatic measuring system attached to the KSM type microscope which registered angles and coordinates of tracks on line. The distribution of azimuthal angles $\psi$ is given in fig. 1 showing a uniform shape that proves that the measurements were correct.

The relativistic particles with $Z = 1$ were separated in fragments ($\theta < \theta_{\lim}$) and newly created particles, mainly mesons ($\theta \geq \theta_{\lim}$) where $\theta_{\lim}$ was calculated according to the formula:

$$\sin(\theta_{\lim}) = \frac{p_f}{p_0}$$

where $p_f$ is the Fermi impulse and $p_0$ is the impulse of the primary nuclei, taking into account the mean loss of energy of iron tracks in emulsion. In our case $\theta_{lim}$ is 15 degrees. All tracks with $\theta < \theta_{\lim}$ were considered protons, deuterium and tritium fragments.

## 3. DETAILS OF THE EXPERIMENT

Nuclear emulsion is a composite medium composed of hydrogen (H, $A_t = 1$), light (CNO, $A_t = 14$), and heavy (AgBr, $A_t = 94$) nuclei. Certainly, there are also other nuclei in emulsions, but their concentrations are too small to be taken into account [9].

The separation technique, which has been used here, is based on the number $n_h$ of target fragments. Events with $n_h = 0$-1 are mainly peripheral $^{56}$Fe-H interactions (interactions with free and quasi-free nucleons) and interactions with other targets (interactions with only one bound nucleon in CNO or AgBr target nuclei). The events having $2 \le n_h \le 7$ are mostly interactions with CNO targets with some admixture of peripheral $^{56}$Fe-AgBr interactions. All events with $n_h \ge 8$ are only due to $^{56}$Fe-AgBr interactions. Following references from [2, 10, 11], this classification of the group of events with $n_h$ = 0−1, 2–7, and $\ge$ 8 can be used to elucidate the nature of the interactions with the three components H, CNO and AgBr of the target emulsion nuclei, respectively.

We present the picture of an interaction Fe-H (Fig. 2 a ) with $n_h$ = 0, and an interaction Fe-AgBr with $n_h$ = 24 (Fig. 2 b ). In Fig.2 the arrows point to a heavy fragment of charge Z = 22, one alpha particles and several shower particles. In Fig.2 b no fragment with Z $\ge$ 2 is produced.

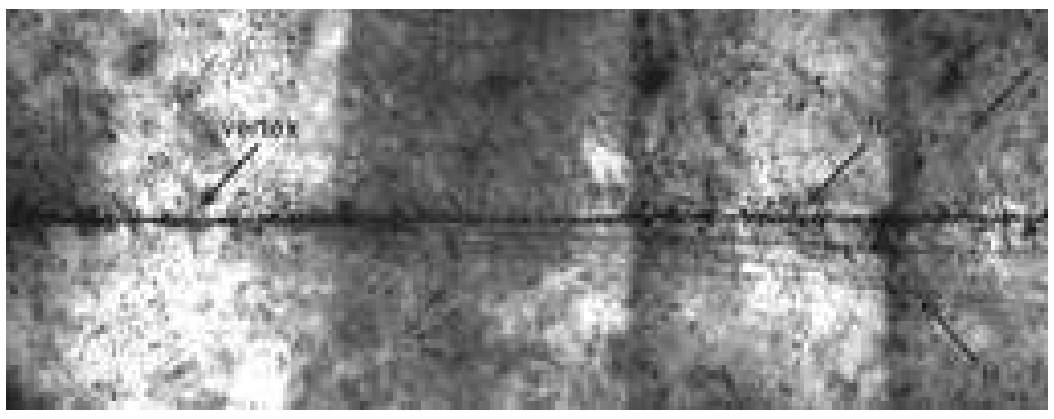

(a)

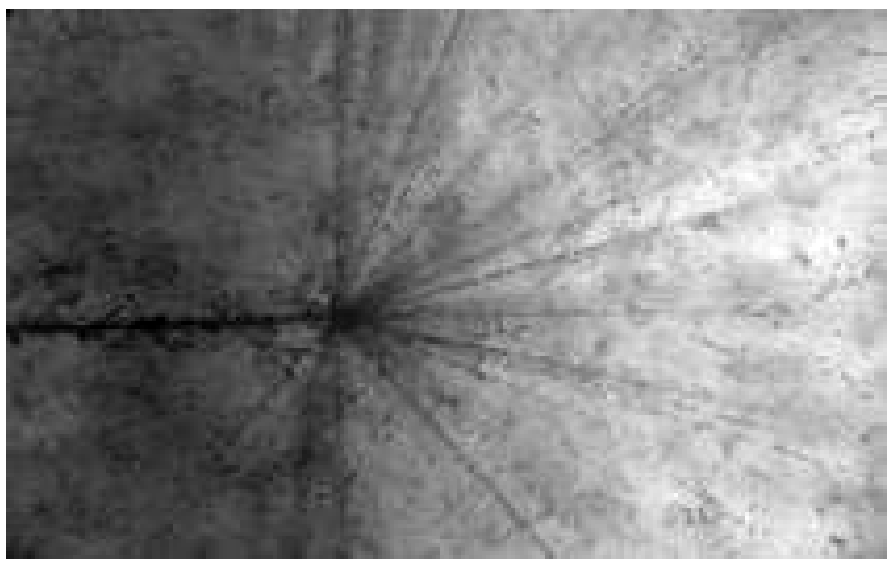

(b)

Fig. [2] - The picture of a peripheral Fe–H collision (a) and a central Fe-AgBr collision (b)

On the basis of the above criteria we obtained 89 H events, 129 CNO events and 165 AgBr events in our sample of 383 events. From 88 events satisfying the necessary criteria for Fe-H interactions in emulsion, we were able to measure the angles in 59 interactions with a favourable geometrical position in the emulsion sheet. We have excluded 19 two-prong stars, one prong being the projectile with unchanged ionization and the second one being the black track of a recoil proton, satisfying the kinematics of elastic Fe-p scattering. The results of the separation along with a comparison with experiments with other beams are given in table 2.

Table 2 shows the percentage of H, CNO and AgBr events from several other experiments as well as from the present one. We observe that the percentage of occurrence of H events increases with the increase of the mass of the beam, whereas the percentage of occurrence of AgBr events decreases with the increase of the mass of the beam.

| Projectile | Energy | Target | | | |
|---|---|---|---|---|---|
| | A GeV | H | CNO | AgBr | Ref |
| $^{14}$N | 2.1 | $12.7 \pm 1.2$ | $32.9 \pm 2.0$ | $54.0 \pm 3.0$ | [12] |
| $^{16}$O | 2.0 | $10.8 \pm 2.0$ | $37.9 \pm 6.0$ | 51.3 | [13] |
| $^{16}$O | 2 | 13.0 | 29.0 | 58 | [14] |
| $^{40}$Ar | 1.8 | $17.8 \pm 1.5$ | $34.6 \pm 1.8$ | $47.5 \pm 3.0$ | [7] |
| $^{56}$Fe | 1.8 | $16.6 \pm 0.8$ | $36.6 \pm 1.7$ | $47.8 \pm 2.6$ | [3] |
| $^{56}$Fe | 1 | $18.27 \pm 2.1$ | $33.68 \pm 2.21$ | $43.08 \pm 2.14$ | Present work |

Table [2] - Percentage occurrence of interactions with different targets

In the CNO and AgBr collision we measured only the angles of the relativistic tracks with $Z = 1$ in order to find those which were emitted in the forward cone, and which were supposed to be fragments of the primary nucleus.

| Charge of fragments Z | $\langle n_Z \rangle$ | | | Ref |
|---|---|---|---|---|
| | Fe-H | Fe-CNO | Fe-AgBr | |
| 1 | $2.52 \pm 0.29$ | $3.19 \pm 0.28$ | $4.94 \pm 0.44$ | Present work |
| | $2.35 \pm 0.17$ | $3.00 \pm 0.13$ | $3.03 \pm 0.09$ | [2] |
| 2 | $0.52 \pm 0.06$ | $1.12 \pm 0.09$ | $1.77 \pm 0.16$ | Present work |
| | $1.17 \pm 0.11$ | $1.61 \pm 0.10$ | $1.62 \pm 0.05$ | [2] |

Table [3] - Multiplicities of projectile fragments in interactions of relativistic $^{56}$Fe nuclei in emulsion

In table 3 we present data on the average multiplicity of fragments with charges Z=1, and Z=2 in interactions in Fe-emulsion, compared with the results obtained in [5], where the primary energy was higher (1,7 A GeV/nucleon). We observe a lower value of the mean number of alpha particles in Fe-H and Fe-CNO collision. The mean values for Fe-AgBr collisions are similar.

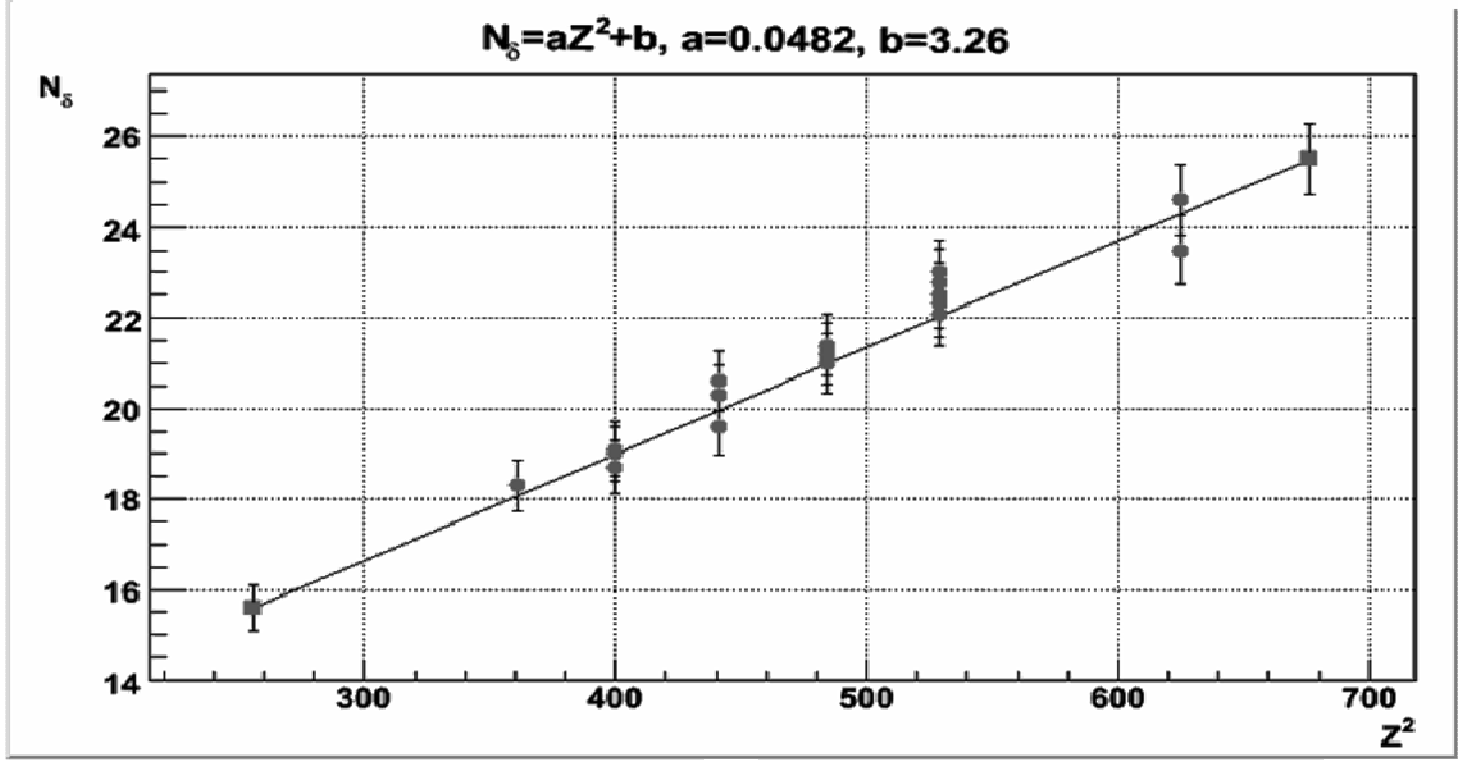

Fig. [3] - Distribution of the density of $\delta$ electrons for fragments with $Z > 10$

The charges of particles with $Z \geq 3$ were mainly evaluated by the density of delta electrons. The light nuclei were identified by gap counting, using a gap length greater than 1 or 2 microns. For the calibration we used the density of delta rays on iron tracks and on a Z=16 heavy secondary, identified by the type of particles in the fragmentation cone. (8 particles of charge one and angles less then 15 degrees and one He particle, assuming that all single charged particles were protons).

In fig. 3 we present the distribution of the measured density $N_\delta$ of delta rays. The dependence of $N_\delta$ of $Z^2$ is expressed by the linear equation: $N_\delta = aZ^2 + b$. The estimated values of the parameters a and b are: $a = 0.0482$, $b = 3.26$.

## 4. DATA ANALYSIS

For the analysis we selected Fe-H - like interactions, in which $n_h$=0,1 (42 stars).

In fig. 4 we present the charge spectrum of secondary fragments emitted from the stars in which no relativistic track of charge one was emitted in the outward cone. We observe the same shape as in other works [15], mainly an abundance of projectile fragments with charge $Z = 1$ and $Z = 2$. We observed only 3 projectiles where $Z = 3$, and then there is a big gap until the charge value $Z = 14$ is reached.

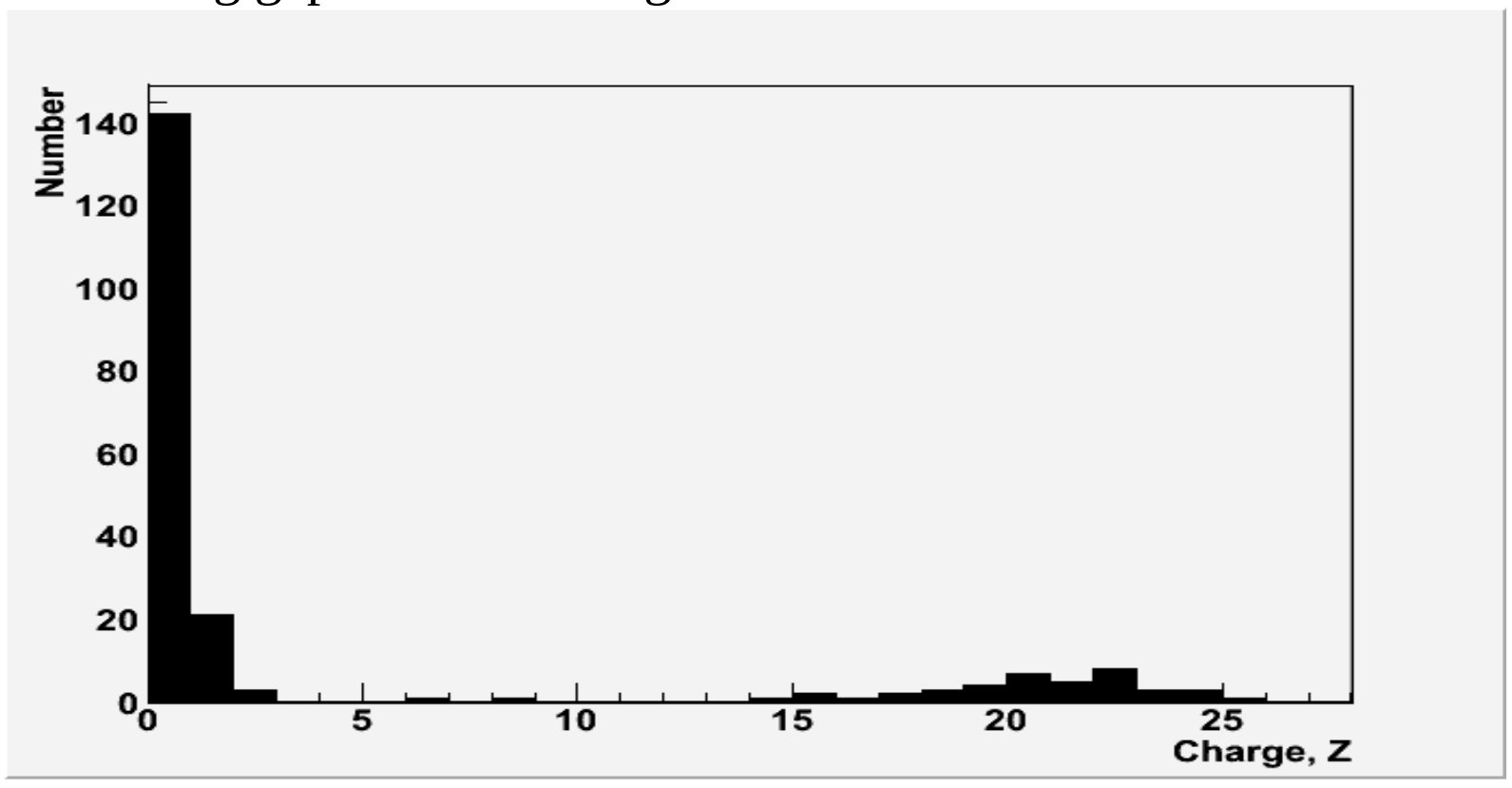

Fig. [4] - Charge spectrum of fragments for H events

As it can be seen there are only 3 events with two fragments. In each case one of the fragment has the value Z=3

The charge topology of the Fe-H stars is given in Table 4:

| Charge values of fragments Fe-H | | | | | | | | | | | | | | | | | | Nr |
|---|---|---|---|---|---|---|---|---|---|---|---|---|---|---|---|---|---|---|
| 26 | 25 | 24 | 23 | 22 | 21 | 20 | 19 | 18 | 17 | 16 | 15 | ... | 9 | 7... | 3 | 2 | 1 | |
| 1 | | | | | | | | | | | | | | | | | | 1 |
| | 1 | | | | | | | | | | | | | | | | 1 | 2 |
| | 1 | | | | | | | | | | | | | | | | 2 | 2 |
| | | 1 | | | | | | | | | | | | | | | 1 | 1 |
| | | 1 | | | | | | | | | | | | | 1 | 1 | 4 | 1 |
| | | | 1 | | | | | | | | | | | | 1 | 1 | 4 | 1 |
| | | | 1 | | | | | | | | | | | | | 1 | 1 | 2 |
| | | | 1 | | | | | | | | | | | | | | 4 | 2 |
| | | | 1 | | | | | | | | | | | | | | 3 | 3 |
| | | | | 1 | | | | | | | | | | | | 1 | 2 | 2 |
| | | | | 1 | | | | | | | | | | | | | 3 | 1 |
| | | | | 1 | | | | | | | | | | | | | 4 | 1 |
| | | | | 1 | | | | | | | | | | | | | 5 | 1 |
| | | | | | 1 | | | | | | | | | | 1 | | 2 | 1 |
| | | | | | 1 | | | | | | | | | | | | 6 | 1 |
| | | | | | 1 | | | | | | | | | | | | 5 | 1 |
| | | | | | 1 | | | | | | | | | | | | 4 | 1 |
| | | | | | 1 | | | | | | | | | | | 1 | | 1 |
| | | | | | 1 | | | | | | | | | | | 2 | 1 | 1 |
| | | | | | 1 | | | | | | | | | | | 1 | 3 | 1 |
| | | | | | | 1 | | | | | | | | | | 1 | 2 | 1 |
| | | | | | | 1 | | | | | | | | | | | 4 | 3 |
| | | | | | | | 1 | | | | | | | | | | 3 | 1 |
| | | | | | | | 1 | | | | | | | | | 1 | 3 | 2 |
| | | | | | | | | 1 | | | | | | | | | 8 | 1 |
| | | | | | | | | 1 | | | | | | | | 1 | 5 | 1 |
| | | | | | | | | | 1 | | | | | | | 1 | 2 | 1 |
| | | | | | | | | | | 1 | | | | | | 1 | 8 | 1 |
| | | | | | | | | | | 1 | | | | | | | 5 | 1 |
| | | | | | | | | | | | 1 | | | | | | 4 | 1 |
| | | | | | | | | | | | | | 1 | | | 1 | 6 | 1 |
| | | | | | | | | | | | | | | 1 | | 4 | 10 | 1 |

Table [4]: The charge topology of the Fe-H stars

In fig. 5 we present the multiplicity distribution of fragments with different charges.

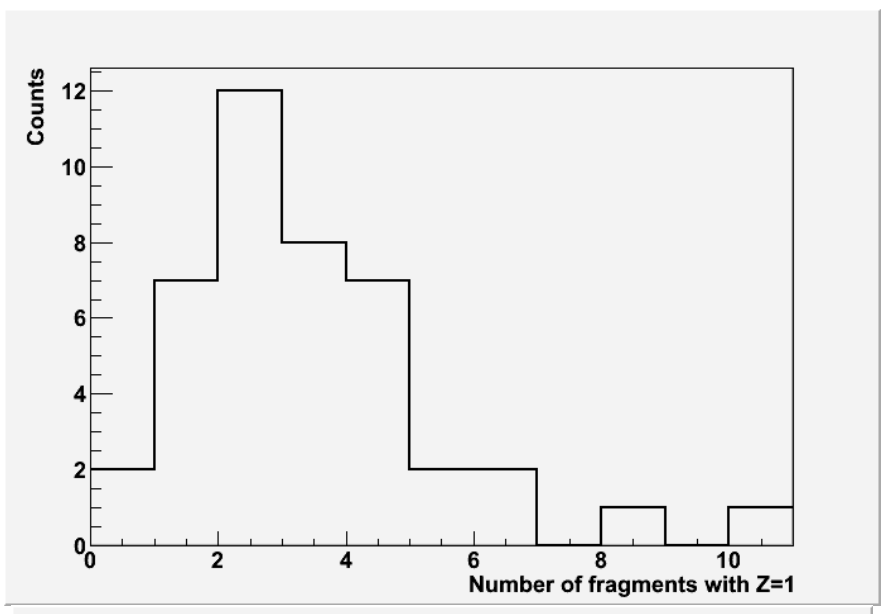

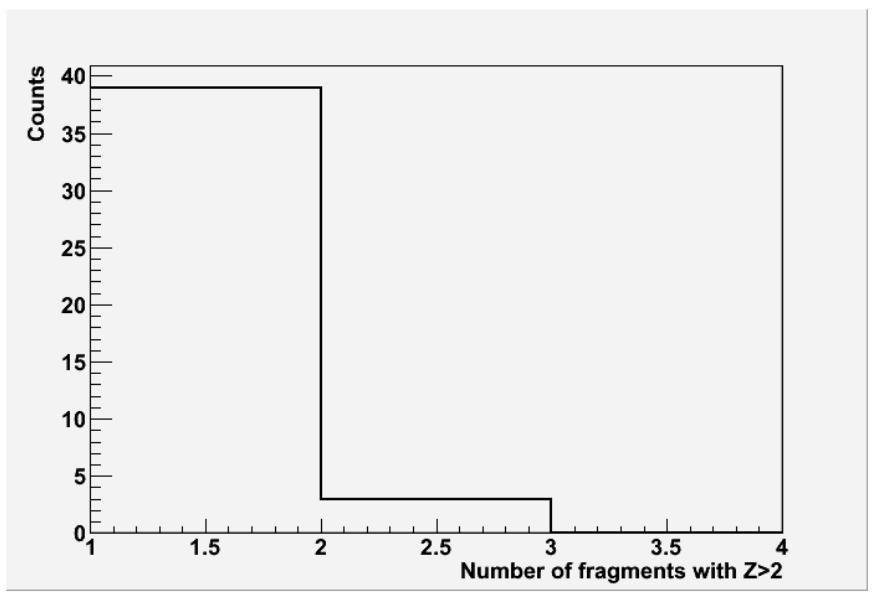

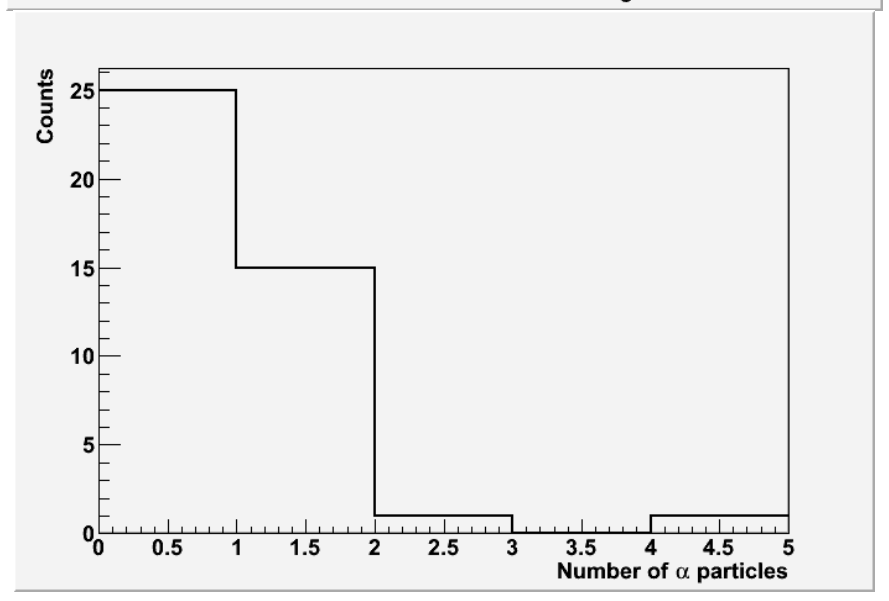

Fig. [5] – Multiplicity distributions for H-type stars: for fragments with Z=1 (upper-left), for fragments with Z>2 (upper-right) and for fragments with Z=2 (lower-left)

In figure 6 and 7 we show the distribution of polar angle distribution and transverse momentum for the fragments with charge Z=1, 2 and for the fragments with Z ≥ 3. Their mean value is given in table 5.

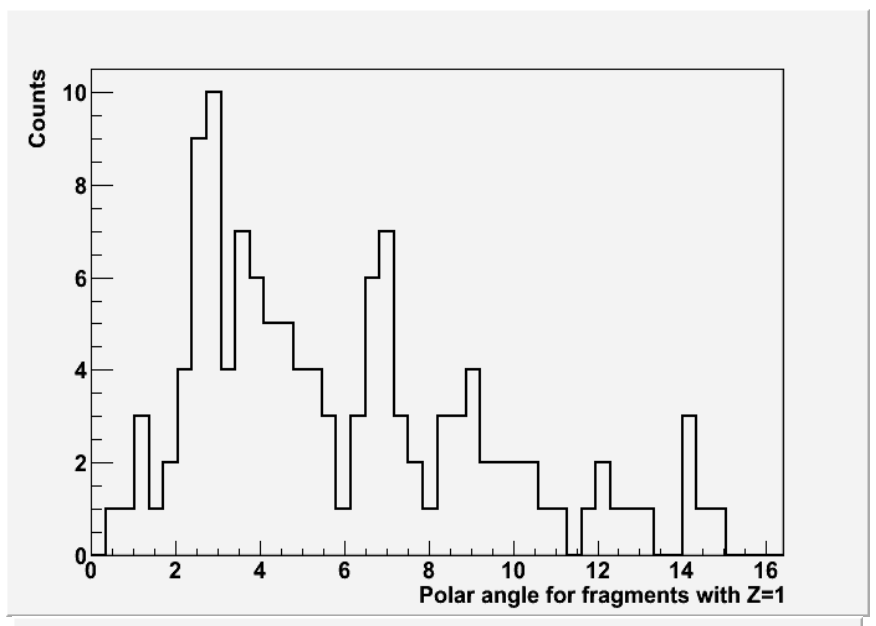

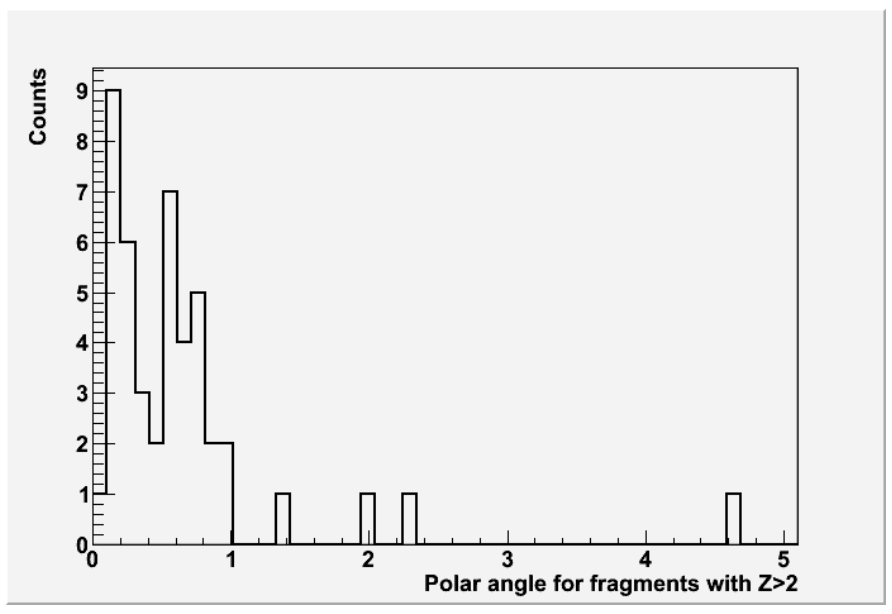

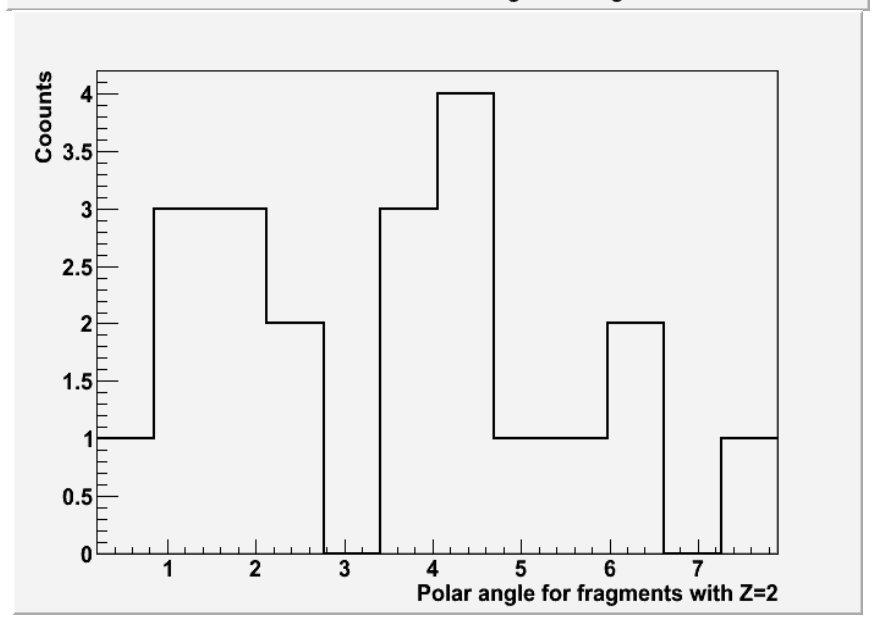

Fig. [6] – Distributions of polar angle for H-type stars: for fragments with Z=1 (upper-left), for fragments with Z>2 (upper-right) and for fragments with Z=2 (lower-left)

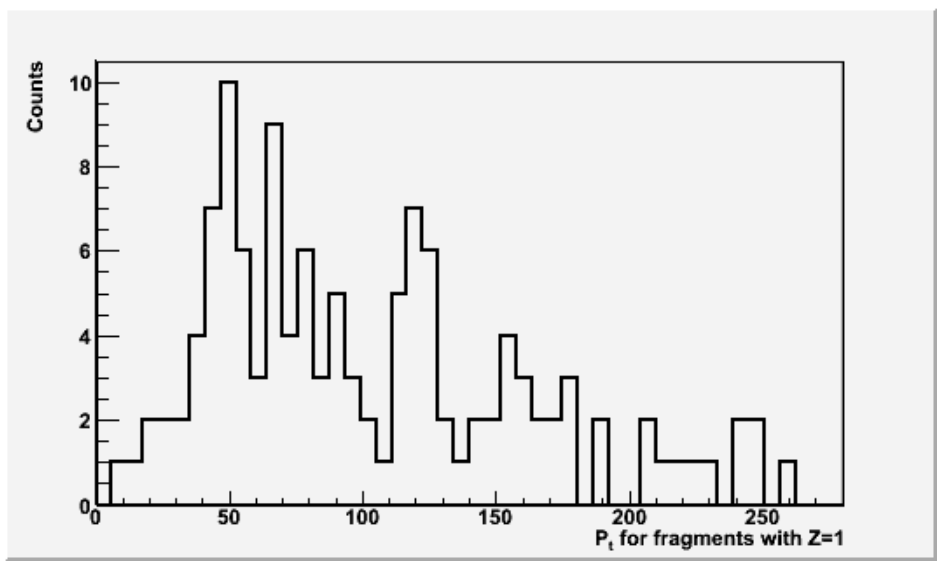

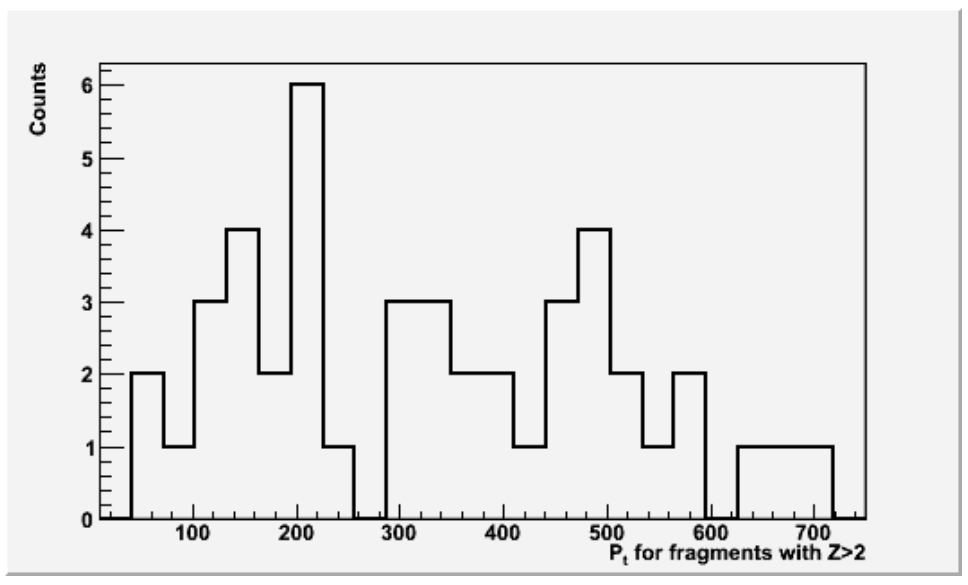

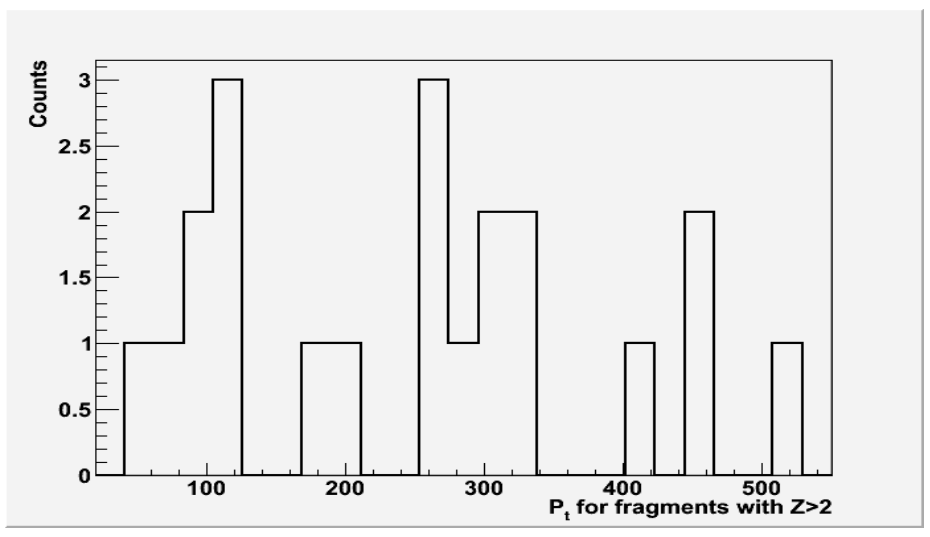

Fig. [7] – Distributions of transverse momentum for H-type stars: for fragments with Z=1 (upper-left), for fragments with Z>2 (upper-right) and for fragments with Z=2 (lower-left)

| | Nr. tracks | $\langle\theta\rangle$ | $<P_t>$ |
|---|---|---|---|
| Z=1 | 123 | 5.97 $\pm$ 0.30 | 103 $\pm$ 5.34 |
| Z=2 | 21 | 3.54 $\pm$ 0.42 | 246.7 $\pm$ 30 |
| Z$\geq$ 3 | 45 | 0.65 $\pm$ 0.11 | 334.5 $\pm$ 26.20 |

Table [5] - Mean value of polar angle and transverse momentum for fragments of charge Z

The various configurations of relativistic fragments produced in peripheral collision in which no target fragments and no mesons are produced reflect the structure of nuclei. As nuclear emulsion provides a complete monitoring of relativistic fragments with an excellent angular resolution, this approach was successfully applied in the Becquerel collaboration [1].

The criteria are the following:

1. No track from the target ( white stars, WS , $n_h$=0 )
2. No meson is produced ( no shower particle in the outward cone $n_s$=0)
3. The charge of the projectile is equal to the sum of charged fragments.

As the percentage of events selected by these criteria is usually very low (4.17 %), in this work we try to find a solution to enlarge this proportion. We suggest the selection of all stars that satisfy the criteria 2 and 3. These are hereafter called ED stars. They represent 11% of our sample of 383 stars.

It follows that the selected stars that can be used for the study of electromagnetic events represent 11,23% of the sample.

| | $<n_{Z=1}>$ | $\langle\theta\rangle_{Z=1}$ |
|---|---|---|
| ED (nh=0) | 3.2 $\pm$ 0.46 | 5.51 $\pm$ 0.47 |
| ED (nh>0) | 2.71 $\pm$ 0.43 | 6.29 $\pm$ 0.48 |

Table [6] - The mean values of the number of shower fragments and the mean polar angle of shower fragments for two categories of events

The reliability of this approach can be checked in different ways. In Table 6 we provide the mean values of the number of shower fragments and the mean polar angle of shower fragments for two categories of events, namely the classical approach ( criteria 1, 2, and 3 ) and the approach proposed in this article (criteria 2 and 3)

Bearing in mind the margin of error we can conclude that the values are in very good agreement.

A more complex study of the properties of the events using the selected stars is under way and will be published in a forthcoming paper.

## 5. CONCLUSIONS

1. The value of the theoretical cross section for the interaction of Fe nuclei in emulsion is in agreement with the observed value in this work.
2. We evaluate the percentage of occurrence of H, CNO and AgBr events as a function of the projectile mass. We observe an increase of this value with the increase in the value of the projectile mass which is in good agreement with results from other works.
3. All of the interactions which have been analysed contained a heavy fragment. The light fragments (Z = 3) were emitted only in those interactions where 2 fragments with Z $\geq 3$ were present.
4. The distribution of the mean number of particles of different charges, of their polar angle and transverse momentum is provided for all the stars attributed to the peripheral interactions.
5. A new set of selection criteria is suggested in order to select electromagnetic interactions. And using it leads to increased number of events of this type as compared to the classical approach

## Acknowledgement

This work was financially supported by the Romanian Ministry of education and Research, under PN II – P4, No 81-044.

## References

[1] Web site of the Becquerel Project http:// becquerel.jinr.ru

[2] G.N. Chernov, K.G. Gulamov, U.G. Gulyamov, V.Sh. Navotny, N.V. Petrov, L.N. Svechnikova, *Fragmentation of relativistic $^{56}$Fe nuclei in emulsion*, Nuclear Physics A412 (1984) 534-550

[3] V.E. Dudkin et al, *Multiplicities of secondaries in interactions of 1.8 GeV/nucleon $^{56}$Fe nuclei with photoemulsion and the cascade evaporations model* Nucl. Phys A 1990 509 783

[4] L.K. Mangotra et al., *Characteristics of $^{56}$Fev-emulsion interactions at 1.7Gev/A* IL Nuovo Cimento 87A, 279 (1985)

[5] V. Singh, B. Bhattacharjee, S. Sengupta, A. Mukhopadhyay, *Estimation of Impact Parameter on event-by-event basis in Nuclear Emulsion Detector*, arXiv:nucl-ex/0412051

[6] Haret C. Rosu, *One mean free path of relativistic heavy ion in nuclear emulsion* Acta Physica Polonica B Vol 25 nr 10 1994

[7] R R Joseph, I D Ojiha, B K Singh and S K Tuli, *Some general properties of projectile fragments in $^{40}$Ar interaction in nuclear emulsion at 1.8 A GeV J Phys G; Nucl Part Phys* 18

[8] W.H. Barkas, *Nuclear Research Emulsion (part I)*, New York and London Academic) p 73, 1963.

[9] M. I. Adamovich et al *Fragmentation and Multifragmentation of 10.6 A GeV Gold Nuclei* ., Eur. Phys. J. A **5**, 429 (1999)

[10] M. El-Nadi, M. S. El-Nagdy, N. Ali-Mossa, A. Abdelsalam, A.M. Abdalla, and A. A. Hamed,, *Fragmentation of $^{28}$Si in nuclear emulsion* J. Phys. G **25**, 1169 (1999)
[11] M.L. Cherry, A. Dabrowska, P. Deines-Jones, R. Holynski, B.S. Nilsen, A. Olszewski, M. Szarska, A. Trzupek, C.J. Waddington, J.P. Wefel, B. Wilczynska, H. Wilczynski, W. Wolter, B. Wosiek, K. Wozniak, *Fragmentation and particle production in interactions of 10.6 GeV/N gold nuclei with hydrogen, light and heavy targets* Eur. Phys. J. C 5, (641)1998
[12] R. Bhanja, N.A.L. Devi, Z.R.R. Joseph, I.D. Ojha, M. Shyan and S.K. Tuli 1983 *Nucl. Phys* A 411 507
[13] B. Jakobsson and R. Kullberg, *Interaction of 2GeV/nucleon $^{16}$O with light and heavy emulsion nuclei Phys*. Scr, 1976 ,13 327
[14] C. Bjarle, N.Y. Herrstom, R. Kullberg, A. Oskarsson and I. Otterlund *The Breakup Of O-16 Into He: An Event By Event Study Of Nucleus-Nucleus Collisions At 75-Mev/A, 175-Mev/A And 2000-Mev/A,* Nucl. Phys. A 381 1982 544-556
[15] G. Singh, K. Sengupta, and P. L. Jain, *Electromagnetic dissociation of $^{32}$S at ultrarelativistic energy in nuclear emulsion*, Phys., Rev. C 41, 999-1004